\documentclass[conference]{IEEEtran}
\IEEEoverridecommandlockouts
\usepackage{cite}
\usepackage{amsmath,amssymb,amsfonts}
\usepackage{algorithmic}
\usepackage{graphicx}
\usepackage{textcomp}
\usepackage{fancyvrb}
\usepackage{xcolor}

\def\BibTeX{{\rm B\kern-.05em{\sc i\kern-.025em b}\kern-.08em
    T\kern-.1667em\lower.7ex\hbox{E}\kern-.125emX}}
\begin{document}

\title{Constellation: An Edge-Based Semantic Runtime System for Internet of Things Applications\\}

\makeatletter
\newcommand{\linebreakand}{%
  \end{@IEEEauthorhalign}
  \hfill\mbox{}\par
  \mbox{}\hfill\begin{@IEEEauthorhalign}
}
\makeatother

\author{\IEEEauthorblockN{Mitch Terrell}
\IEEEauthorblockA{\textit{Distributed Computing Systems Group} \\
\textit{University of Minnesota}\\
Minneapolis, USA \\
terre101@umn.edu}
\and
\IEEEauthorblockN{Yixuan Wang}
\IEEEauthorblockA{\textit{Distributed Computing Systems Group} \\
\textit{University of Minnesota}\\
Minneapolis, USA \\
yixua003@umn.edu}
\and
\IEEEauthorblockN{Bhaargav Sriraman}
\IEEEauthorblockA{\textit{Distributed Computing Systems Group} \\
\textit{University of Minnesota}\\
Minneapolis, USA \\
srira048@umn.edu}
\and
\IEEEauthorblockN{Soumya Agrawal}
\IEEEauthorblockA{\textit{Distributed Computing Systems Group} \\
\textit{University of Minnesota}\\
Minneapolis, USA \\
agraw184@umn.edu}
\and
\IEEEauthorblockN{Matt Dorow}
\IEEEauthorblockA{\textit{Distributed Computing Systems Group} \\
\textit{University of Minnesota}\\
Minneapolis, USA \\
dorow016@umn.edu}
\and
\IEEEauthorblockN{Zachary Leidall}
\IEEEauthorblockA{\textit{Distributed Computing Systems Group} \\
\textit{University of Minnesota}\\
Minneapolis, USA \\
leid0065@umn.edu}
\linebreakand
\IEEEauthorblockN{Abhishek Chandra}
\IEEEauthorblockA{\textit{Distributed Computing Systems Group} \\
\textit{University of Minnesota}\\
Minneapolis, USA \\
chandra@umn.edu}
\and
\IEEEauthorblockN{Jon Weissman}
\IEEEauthorblockA{\textit{Distributed Computing Systems Group} \\
\textit{University of Minnesota}\\
Minneapolis, USA \\
weiss039@umn.edu}
}

\maketitle

\begin{abstract}
With the global Internet of Things (IoT) market size predicted to grow to over 1 trillion dollars in the next 5 years, many large corporations are scrambling to solidify their product line as the de-facto device suite for consumers. This has led to each corporation developing their devices in a siloed environment with unique protocols and runtime frameworks that explicitly exclude the ability to work with the competition’s devices. This development silo has created problems with programming complexity for application developers as well as concurrency and scalability limitations for applications that involve a network of IoT devices. The Constellation project is a distributed IoT run-time system that attempts to address these challenges by creating an operating system layer that decouples applications from devices. This layer provides mechanisms designed to allow applications to interface with an underlying substrate of IoT devices while abstracting away the complexities of application concurrency, device interoperability, and system scalability. This paper provides an overview of the Constellation system as well as details four new project expansions to improve system scalability.
\end{abstract}

\begin{IEEEkeywords}
Internet of Things, Edge computing
\end{IEEEkeywords}

\section{Introduction}
The Internet of Things (IoT) is the interconnection of the physical and digital world utilizing sensors and actuators to automate, personalize, and otherwise increase the efficiencies of everyday tasks. This new paradigm introduces endless applications for enhancing the daily lives of ordinary people as well as mechanical processes in industry. Although the devices used in IoT applications are heterogeneous and have disparity in bandwidth, connectivity, etc. Constellation is an Edge-based Semantic Runtime System for Internet of Things Applications which deals with the disparity and dynamism of the network.
The goal of our system is to (a) utilize 'semantic equivalency' of devices and subsystems in an environment to avoid conflict and increase efficiencies, (b) develop new ways of resolving conflicts that arise dynamically in a local environment and (c) provide a simple yet robust means for developers to rapidly develop IoT applications.\par
It is built with various policies and other capabilities to resolve disparity and counter the dynamism in the network.\par
To handle scalability and developer robustness in the IoT world an IoT system must rapidly be adapting as new devices and new protocols are developed. To handle this, the our system provides the ability to interact with the open source WebThings project as a secondary mediator for devices. The WebThings project allowed for access to hundreds of industry device drivers under a unified protocol following the W3C standard. By emulating virtual devices running on a WebThings Gateway, our system was able to connect with the many industry devices supported by WebThings through their diverse protocol suite such as  Zigbee, Zwave, Bluetooth, Apple HomeKit, Weave and IP. In addition the ability to offload device communication overhead to the WebThings Gateway opened many doors in terms of scalability.\par
Edge nodes utilize a local cache for IoT device queries to improve scalability by attempting to reduce the amount of device querying required. Each edge node stores and models recent device query results to be able to predict future queries without communicating with the device. Resolving queries through cache prediction can provide a significant reduction in querying at the expense of some amount of query result prediction error. This trade-off between query reduction and acceptable error can be tuned to the application developer's desire.\par
Privacy is a rising concern in the IoT environment. Coordinating with hundreds of applications and clients expose Constellation to threats and attacks. One of the main concerns is data leakage caused by untrustworthy applications and service providers. In this study, we demonstrated the critical security and privacy requirements of Constellation and introduced corresponding defenses mechanisms to secure private data.\par
In terms of architecture, Constellation is envisioned as a system with IoT devices distributed across different locations like in a smart city. These IoT devices communicate with the edge nodes that are organized in a peer to peer network. One of the important problems in the peer to peer network is Fault Tolerance. In this work, we have implemented strategies that have enabled clients to be transparent to edge node failures and IoT device failures.

\section{System Overview}
In this section, we provide an overview of the system architecture as well as a look at some of the existing constructs built into the system for simplifying the management of devices and optimizing the execution of concurrent applications.

\subsection*{Requirements}
We aim to provide a system designed for managing and communicating with heavily dispersed IoT devices capable of carrying out both sense and actuation requests. Applications utilizing these devices should be able to run and communicate with the system from the edge or in the cloud.\par
Beyond managing and communicating with devices, the system should also aim to provide optimizations via a software layer between applications and the IoT devices. System optimizations should provide improved scalability while still maintaining flexibility to support a variety of heterogeneous devices and application goals.\par
Constellation was designed to meet these requirements by introducing a distributed operating system layer between applications and devices to ease application development and optimize overall system performance.

\subsection*{Architecture}
Our system is organized into a peer-to-peer network of edge nodes. Each edge node is responsible for managing and communicating with the IoT devices in its vicinity. Applications make requests to an edge node which then carries out the request on the underlying IoT device(s). An example organization of devices in the Constellation network is shown in Fig. \ref{p2p_network}. This design offers scalability as well as a level of redundancy and fault-tolerance as nodes leave and join the network.\par

\begin{figure}[htbp]
\centerline{\includegraphics[width=\linewidth]{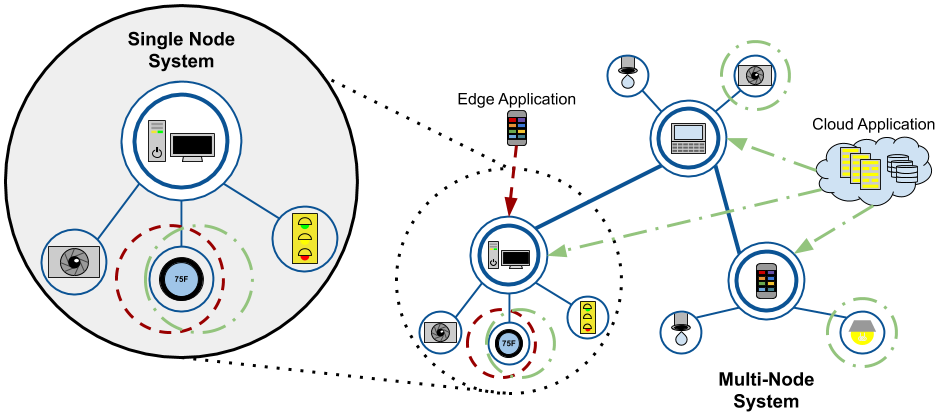}}
\caption{System Peer-to-Peer Network}
\label{p2p_network}
\end{figure}

Each node in the network is comprised of three software layers, two of which provide a universal interface layer to applications and devices respectively, and a middle layer where task execution and system maintenance are performed. This organization is shown in Fig. \ref{node_architecture}.\par

\begin{figure}[htbp]
\centerline{\includegraphics[scale=0.8]{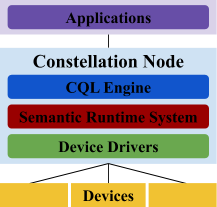}}
\caption{Node Architecture}
\label{node_architecture}
\end{figure}

A node's top layer, the CQL Engine, interfaces with external applications using a declarative query interface inspired by jQuery and SQL. This layer provides a simple interface for IoT applications to communicate with a vast array of potentially disparate devices in a portable way.\par
The real-time execution engine comprises the second layer of the Constellation node stack. This layer performs the execution of IoT tasks, maintains state and connectivity to the underlying IoT devices, and handles maintenance of the peer-to-peer network. This layer is also responsible for recognizing the synergies and conflicts inherent when enabling many concurrent applications to operate on the same collection of physical devices.\par
The lowest node layer is a device driver layer that enables Constellation to communicate with potentially any device. Each driver is envisioned to be developed by the device manufacturer, similar to how drivers are developed in a traditional operating system, to make the device compatible with the system.\par
All of these components provide the means for Constellation to be a robust and efficient system with the ability to handle numerous concurrent devices and applications.

\subsection*{Constellation Query Language (CQL)}
Constellation query interface layer enables IoT applications to interface with underlying IoT devices without requiring the applications to account for the individual requirements of each IoT device, even those from different manufacturers or using different networking technologies. The language itself resembles that of english, like SQL, and enables the visualization of IoT objects as simple but dynamic sets which are formed based on criteria specified by the user, like in jQuery. The idea behind CQL was that webpages are often used as a virtual version of storefronts or other shared spaces, so why not take a technology like jQuery which allows for elements on a webpage to be searched based on their attributes, and apply that same idea back to physical spaces?

The language is used to specify common IoT tasks such as sensing, actuation, events, and device discovery. Tasks can be periodic or aperiodic and can have real-time deadlines enforced. The language is composed of four essential statements, SENSE, ACTUATE, EVENT, and FIND. The first three statements cover the three types of common IoT actions recognized by the World Wide Web Consortium (W3C) as fundamental IoT operations. The four core Constellation query commands are outlined in Tab. \ref{command_tbl}.

\begin{table}[htbp]
\caption{System Query Language}
\begin{center}
\begin{tabular}{|c|c|}
\hline
\textbf{Query}&\multicolumn{1}{|c|}{\textbf{Purpose}} \\
\hline
FIND & Discover IoT devices in the area \\
\hline
SENSE & Query a device for environment information \\
\hline
ACTUATE & Query a device to perform an action \\
\hline
EVENT & Simple ‘if-this-then-that’ logic\\
\hline
\end{tabular}
\label{command_tbl}
\end{center}
\end{table}

\par The FIND statement is the most fundamental statement as it facilitates the discovery of available IoT resources by an application.  The FIND statement is a statement essential for IoT applications to locate IoT devices that suit their requirements so that subsequent statements of the other types can be performed on said devices. The SENSE statement is used for applications to gather sensor readings from a set of devices returned by a FIND statement. The statement can be periodic, possess a real-time deadline, and can also specify a tolerance level for the sensor reading’s accuracy which can be used to enable optimizations such as caching, where the device  does not always have to be the source of the reading, saving energy and latency. The ACTUATE statement, like the SENSE statement, operates on a set of devices as returned by the FIND statement. It can likewise have a cardinality, period and/or deadline, but is distinguished from the SENSE statement by a lack of a DELTA parameter as well as by the existence of the PARAMS modifier. Finally, EVENTS in the system represent simple ‘if-this-then-that’ logic which is executed atomically by the compute nodes, rather than requiring the network latency of communicating with the client application. The events can be either ‘triggered’ e.g. by a sensor reading exceeding a particular threshold, or they can simply be periodic, executing at each epoch defined by a static period.

\subsection*{Device Sets and Device Driver}
The concept of device sets(DevSet) is a key innovation of Constellation. A device set is a set of IoT devices that can be considered functionally equivalent for an application. Devices in the same DevSet are able to be interchangeably deployed. Thus, it provides the flexibility to dynamically prioritize different devices based on criteria their ability to meet real-time deadlines, conserve energy, provide accurate results, etc. This capability also enables optimization for energy efficiency, accuracy, and latency, etc. In addition, this also provides fault tolerance by utilizing the functionally equivalent device when failures happen. \par
Constellation enables interoperability between devices across manufacturers through the use of a device driver interface. The interface makes for the system as a whole while being transparent to the end-user. This java-based interface contains the driver file that is the device driver and performs functionality, and an XML-based ‘device manifest’ file that describes the device’s capabilities and the contents of its driver/API. Constellation also supports the WebThings driver interface, which unlocks the ability for hundreds of devices that are compliant with the WebThings Gateway.

\subsection*{Run-time Engine}
As can be seen in Fig. \ref{node_architecture}, the Semantic Runtime Engine lies between CQL Engine and Device Drivers. The Semantic Runtime engine consumes CQL query and converts it into objects. The object contains various information like cardinality, start time, URI of the IoT device (actuator or sensor). In the system, these objects are called 'tasks' defined as the combination of requests sent to IoT devices, so if multiple requests are performed in a single query, all these requests are performed as sub task that can be executed in parallel. \\
In this system, we also maintain virtual device objects (Device Drivers in Fig. \ref{node_architecture}).  These device objects contain metadata about the IoT device. For example, location, execution time, API, reference to device drivers etc.

\subsection*{Device Energy Profiles}
Some IoT devices operate on battery power with a limited and scarce battery capacity. To improve battery utilization, Constellation introduces mechanisms to track and reduce energy consumption. This is accomplished by separating devices into three separate categories based on the device's energy constraints and capabilities: non-energy-aware (default), Metered, and Sleepy devices. Non-energy-aware devices do not have any energy constraints and therefore operate under normal conditions.\par
Metered devices are described by the device energy level and a profile modeling the energy consumed by the device's functionalities. Energy profiles can either be static or dynamic models. Static profiles are given by predefined energy consumption values from the device driver's manifest, while dynamic profiles use a multi-factor linear regression model to track the device actions and corresponding device energy depletion rate. This allows the system to track energy constrained IoT devices.\par
Sleepy devices are devices that support a low-power mode that can be entered to conserve energy when the device is not in use. Constellation utilizes sleepy mode by tracking when the device is not currently in use. The system then spawns sleep and wake tasks to manage the device's duty cycle.

\subsection*{Client to Constellation Specification}
This section describes the client-side API that is necessary for interfacing with Constellation. The client-side API allows for easy development of applications that can interact with a node or even a system of nodes. The client side API allows for the connection to the system from a client allowing for the ability to issue queries on the devices connected to the system such as sense and actuation commands. Connecting to the system is very similar to connecting to a database management system like MySQL. All that is required is the socket address of the system node. A connection call which will produce an object through which queries can be sent.  To close the connection, which will cancel all tasks created by this client, simply call the close() function on the connection object. To query the device a query command can be sent to the object returned from the connection command above. This query command has a string as a parameter and will return a result object which indicates the success or failure of the query as well as the any information that was requested. 

\section{Core 1: Fault Tolerance in a Multi-Node Setup} 
The system was envisioned as a multinode setup where several edge nodes are connected and are responsible for a set of IoT devices. This section describes how a multinode setup is created in the system. Fault tolerance is important as it will help clients to interact with IoT devices even after failure of certain edge nodes. We designed a heterogeneous network of edge nodes defined as Leader Node and Edge Node. Leader nodes are edge nodes that are stable and have high compute capacity. And Edges are volatile nodes that may or may not have high compute capacity.
\subsection*{Network Topology}
The edge nodes that are a part of system are volatile and can have varied computational capacity and network bandwidth. As a result, the system experiences a constant threat of node failure. Failover mechanisms have to be put in place to avoid discontinuity in service. Constellation has a hierarchical network topology, inspired by KaZaa, to facilitate fault tolerance. \cite{kazaa} \\
Along with edge nodes and IoT devices, the system also consists of leader nodes that are more powerful and less volatile than the average edge node. A leader communicates with a set of edges while maintaining the metadata about them. Furthermore, the leader performs health checks in its cluster to determine the status of each edge node. If an edge fails, the leader transfers the IoT devices managed by the failed edge to another edge.\\
The leaders are connected within themselves in a tree-like structure. They communicate with each other to either send heartbeats to check for failure, or to propagate queries. When a leader receives a query, it checks within its cluster to serve it. If it cannot, it propagates the query to neighboring leader nodes.\\

\begin{figure}[htbp]
\centerline{\includegraphics[width=\linewidth]{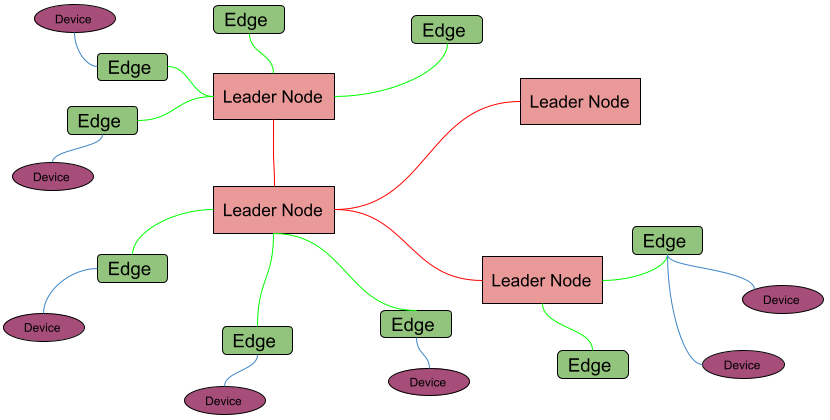}}
\caption{Multi-Node Topology}
\label{topo}
\end{figure}

\subsection*{Discovery}
A new edge that wants to join the system needs a list of leader nodes to find the closest leaders before sending a connection request. A centralized persistent store that stores the leader addresses is used to keep track of the leader nodes in the system. Additionally, each edge node maintains a list of potential leaders that gets periodically updated using this list. The list is updated when a new leader is elected or when an existing leader fails. The store also keeps a list of potential leader addresses. When the systems need a new leader node, typically after a leader failure or when the system is scaling, it utilizes this list. If the centralized store is not reachable, a bootstrap properties file containing the addresses of the initially deployed leader nodes is accessible by the edge node.\\

When a new edge wants to join the system, it first looks up the persistent store that contains the IP of all the leaders. Then it picks the closest leader from the list and sends a connection request to the leader. If the leader can accept another edge node as its neighbor, it sends an acknowledgment and updates its data structures as shown in Fig. \ref{EdgeDiscovery}.\\

\begin{figure}[htbp]
\centerline{\includegraphics[width=\linewidth]{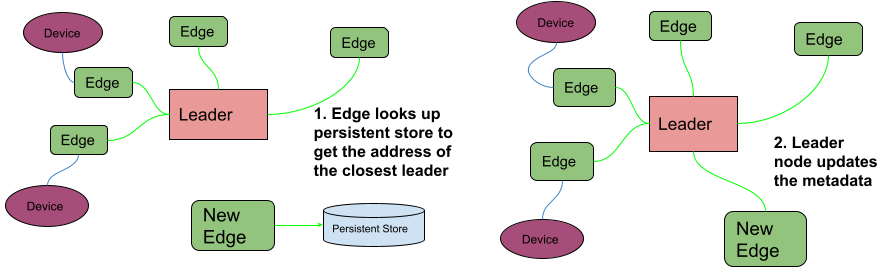}}
\caption{Edge Discovery}
\label{EdgeDiscovery}
\end{figure}

When a node wants to join as a leader, the address of the node is stored as a potential leader. Then, it joins the system as a normal edge node. In the event of a leader failure or scaling, the daemon picks up a valid leader node from the potential leader list and updates the data in the store. This process is depicted in Fig. \ref{LeaderDiscovery}

\begin{figure}[htbp]
\centerline{\includegraphics[width=\linewidth]{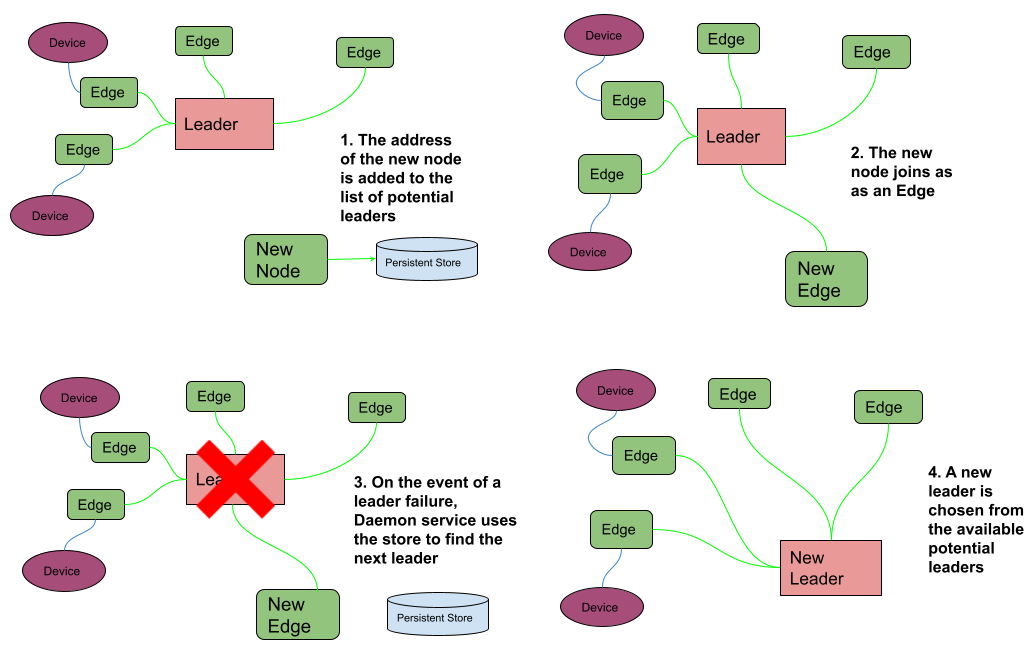}}
\caption{Leader Discovery}
\label{LeaderDiscovery}
\end{figure}

\subsection*{Scalability}
To make the system scalable and stable in cases where new edge nodes are added or removed from the system,  two services are designed that will update the network dynamically.\\
Scaler service: This service will keep track of the number of leaders in Constellation. The number of leaders in the network should be sufficient to handle the edge nodes in the system. Based on the count of edge nodes, this scaler service will determine the number of leaders in the system. In addition, it will update the current number of leaders to a centralized store. \\
Daemon service: This service is responsible for electing a new leader in case the number of leaders in the system network becomes less than determined by the Scaler service.
\subsection*{Handling Failures}
Based on the fault tolerance design, our system is able to handle failures happen in both leader and edge nodes.\\
Leader Failures are handled by the following steps:
\begin{itemize}
\item Heartbeats are established between every leader-edge and every leader-leader connection to immediately detect leader failure. 
\item Each edge maintains a list of potential leaders which is periodically updated. The list is stored along with the RTT time of the packet sent from the edge to that leader. In case its leader fails, the edge can connect to the closest viable leader.
\item Each leader calculated a threshold value, say t, based on the available bandwidth and memory. In order to avoid overloading, it will maintain only up to t number of edge nodes. 
\item Scaler service maintains a minimum number of leaders and tries to scale the network when every leader is next to their threshold. 
\end{itemize}

Edge failures are handled by the following steps:
\begin{itemize}
\item Heartbeats are established between every leader-edge to detect edge failures.
\item Each leader maintains a map of edge and IoT devices, so in case a edge node fails the abandoned IoT devices are failed over to another edge node.
\item As soon as an edge node failure is detected by the leader node, it frees up all the IoT devices, so that when a new edge node discovers the IoT devices, the leader node attaches the device to the new edge node. Leader node then updates its map correspondingly.
\end{itemize}

\section{Core 2: Large Scale Industry Device Support via WebThings}
A core motivation for the Constellation project was to add scalability to IoT device systems. In order to test and experiment with scalability a system needs the ability to interface with a variety of IoT devices without having a massive overhead of device support. Due to the diversity in IoT device types, brands and the wide variety of communication protocols being used, the Constellation project looked to add a method of quickly interfacing with an large stack of diverse IoT devices.
\par The WebThings project is an open source project initially started by Mozilla to create a decentralized internet of things that is intended to unify the application layer of the IoT devices by linking them together using a standardized data model and protocol to make the devices accessible on the web via the W3C standardization. WebThings provides an additional abstraction layer that hides the complexity of working with the many IoT devices' communication protocols such as Zigbee, Zwave, Bluetooth, Apple HomeKit, Weave and IP. We felt that working with the WebThings project seemed like a logical step as WebThing's large open source community has provided device driver support for hundreds of devices that are currently on the market. The WebThings project's easy to use rest API meant that Constellation can interface directly with all of these devices using the the system query language without the overhead of creating individual device drivers for each device that interfaces with Constellation network.

\begin{figure}[htbp]
	\centerline{\includegraphics[width=\linewidth]{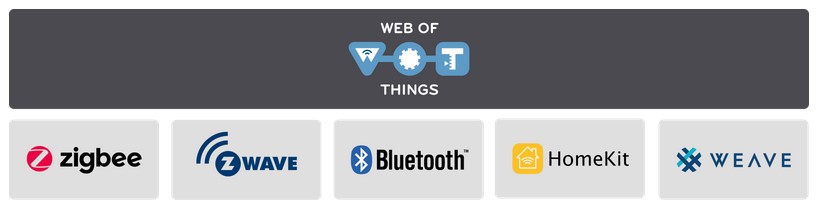}}
	\caption{WebThings application layer}
	\label{webthings_app_layer}
\end{figure}

\subsection*{WebThings Environment Creation}
WebThings works by running a WebThings Gateway in a location near where the devices plan to run. This Gateway acts as a IoT device hub which is able to run the Gateway source code allowing for communication to the devices around it. The Gateway software was initially intended to run on a dedicated Raspberry Pie with Zigbee, Zwave and Bluetooth dongles plugged into the Gateway allowing it to connect to local devices, but the Gateway can also run via Docker containers or on Linux.
\par We started the WebThings project addition by creating a WebThings environment. This included running a Gateway locally in a room on a Raspberry Pie and connecting a number of devices to the Gateway. The Gateway web interface can be seen below in Fig. \ref{webthings_ui}. In addition in Fig. \ref{webthings_smart_strip} an example of the convenient web interface is displayed as an example of how the WebThings project exposes the functionality of a Phillips Hue smart light strip via the WebThings web interface.

\begin{figure}[htbp]
\centerline{\includegraphics[width=\linewidth]{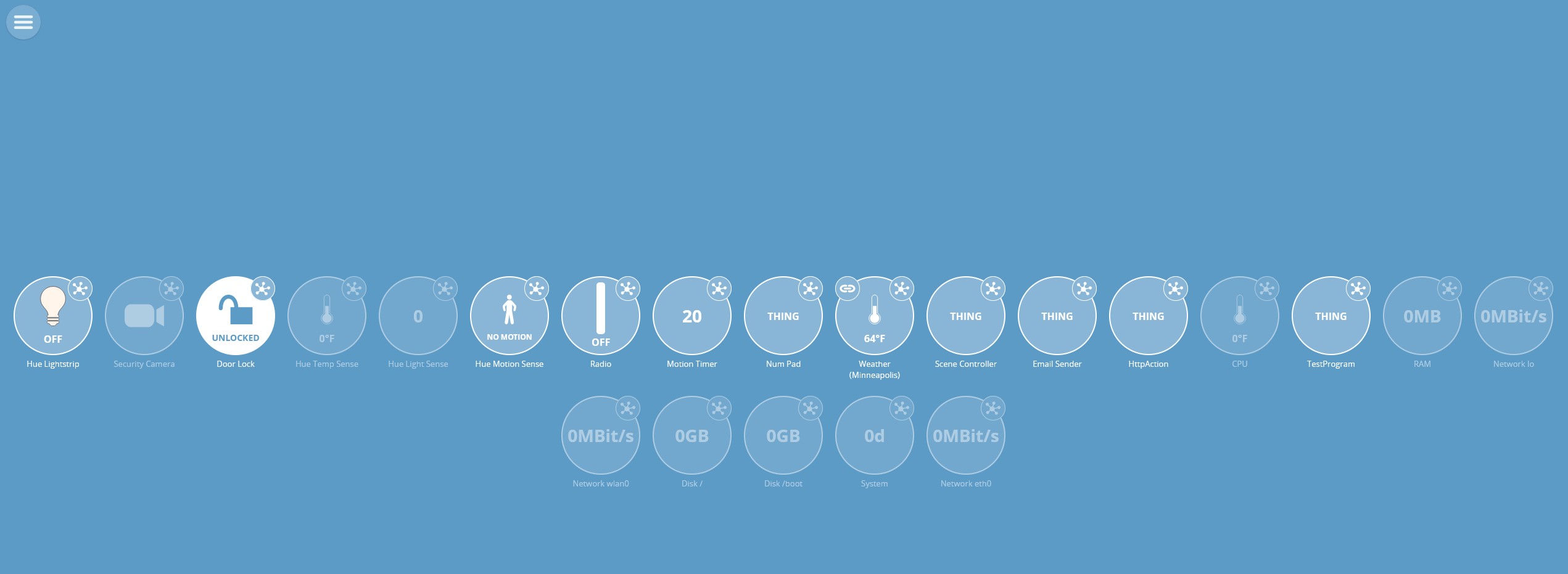}}
\caption{WebThings Gateway web interface}
\label{webthings_ui}
\end{figure}

\begin{figure}[htbp]
\centerline{\includegraphics[width=\linewidth]{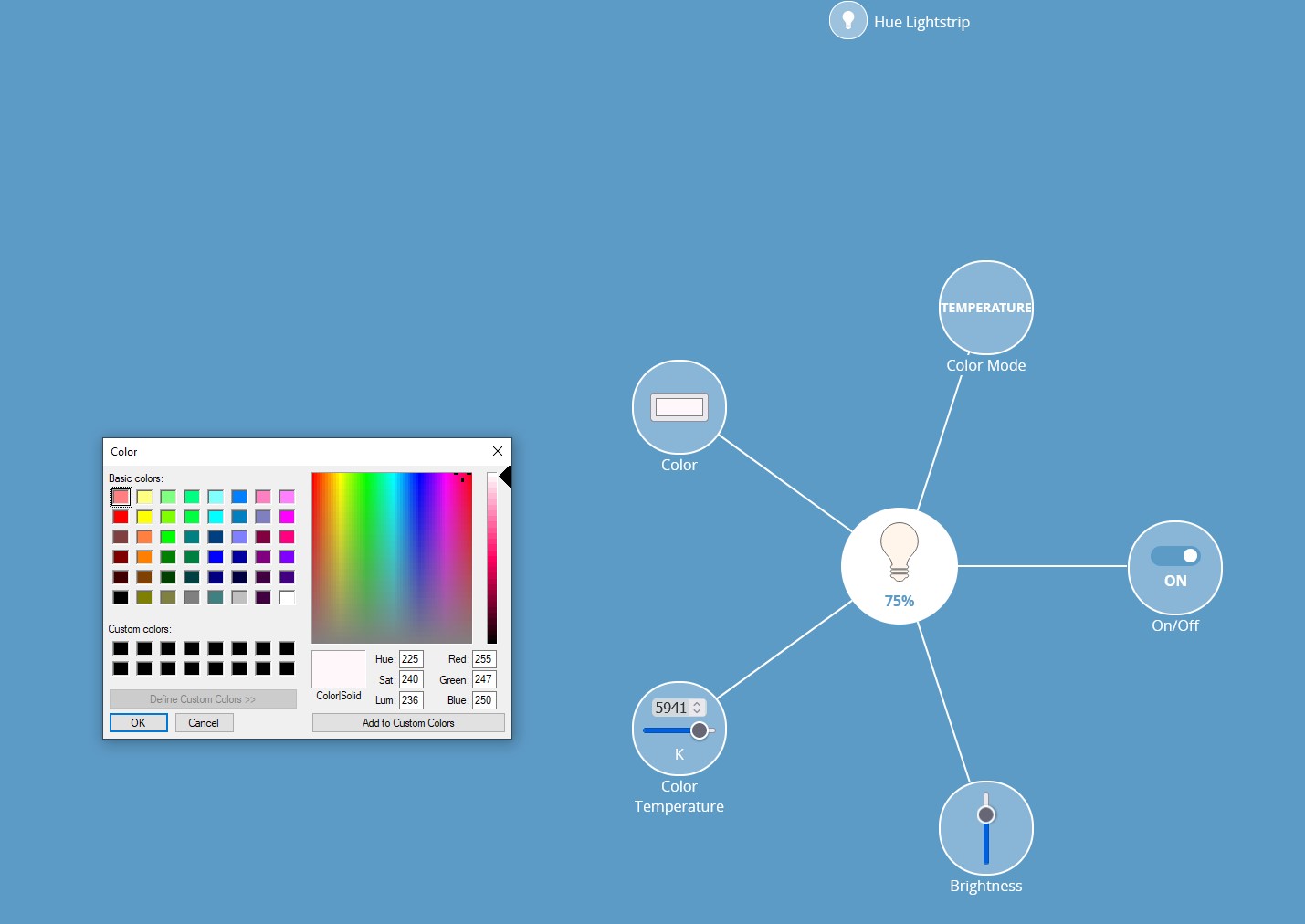}}
\caption{WebThings application layer}
\label{webthings_smart_strip}
\end{figure}

\subsection*{Constellation + WebThings Rest API}
The WebThings project allows for interfacing with a Gateway and its corresponding IoT devices via a rest API utilizing http GET's and PUT's that transfer Json packets back and forth with information about desired action or the state of the environment. Constellation was able to utilize this rest API to leverage the WebThings project as essentially a mediator for IoT device interaction. By connecting to a WebThings Gateway with the rest API, the system was able to query what devices are connected to a Gateway and what functionality is exposed for each of these devices. Once that information was available, internally the system was able to simulate a virtual device that can interact with Constellation query languages the same way that any other device would be able to interact with the network.
\par To provide a clearer understanding of how WebThings and Constellation interact an example of how our system might interact with a device running on a Gateway is provided. An additional suite of commands were added to the CQL was added that allowed for the adding of WebThings Gateway devices. For example, an application developer would pass in a URL to the WebThings Gateway. From there our system would request a device ledger from the Gateway via the rest API for which WebThings will return a Json packet with all devices and their functionality running on that Gateway. Constellation would then parse that packet and create virtual devices internally that would mimic all the functionality of all devices running on the Gateway. As an example, a Phillips Hue light strip might have the properties such as "On/Off", "Color", "Brightness" and it might have actions such as "Turn On/Off" or "Change Color". The system mirrors this functionality internally, so then at the time of receiving a standard CQL command such as an actuation "Turn on Light" it would interact with the device the same way that is described in section 2(iii) except instead of issuing requests directly to the device, our system instead uses the WebThings rest API to request the action or information from the Gateway which in turn requests from the device itself.

\subsection*{Scalability via Devices and Computational Offloading}
Utilizing the WebThings project with Constellation enabled the it to quickly prototype scalability experiments with hundreds of industry standard IoT devices. This removed the dependency on hand made IoT devices and hand written device drivers. Not only did this allow the system to interact with real world IoT devices, but it also added the ability to keep the project modern as new devices come out. This is because the WebThings project is an open source project which has many contributors adding support for new devices all of the time.
\par An unexpected scalability addition can be exposed by using \textbf{many} Gateways running \textbf{many} devices in \textbf{many} locations. By offloading the computational and communication burden of in 'room' device manipulation, the Gateway's free up the system and allow it to handle larger and larger sets of queries. Suddenly a whole new 'room' of devices can be added to the system without a worry of it having too many devices to manage, since the local Gateways (often running on Raspberry Pie's locally) are handling the the actual device manipulation/communication and are just relaying success or failure back to the system.

\section{Core 3: Edge Node Caching}
To carry out SENSE requests, Constellation edge nodes need to constantly query the corresponding IoT devices to retrieve the data requested by applications. As the number of devices and applications using those devices grows, the increased number of requests may put a strain on the IoT devices. Reducing the number of queries handled by devices by offloading these requests to edge nodes can reduce device energy consumption and network usage while also potentially improving latency.\par
Many real-world IoT devices intended for data collection commonly collect data that exhibit stable or predictable data patterns. For example, an outdoor thermometer measures temperature that typically will gradually fluctuate and follow a predictable diurnal pattern of high temperatures during the day and low temperatures at night.\par
To alleviate the amount of device querying required, Constellation introduced device data caches at the edge nodes to collect, model, and predict future requests without re-querying the device. Edge node caches are also tied to each device so multiple applications can utilize a single shared cache, further unburdening the device from duplicate requests that can be satisfied by information stored in the cache.\par
Since IoT devices collect a wide variety of data types and applications require differing levels of acceptable constraints, it is important that the cache offer a flexible configuration that can adapt to the specific IoT device and application use case. The following sections describe the design of the caching mechanism, customization features, and a brief analysis of performance.
\subsection*{Modeling}
To avoid querying IoT devices, the edge node cache must be able to accurately predict the current value that would be returned by the device if it were to be queried. Constellation achieves this by modeling the data points collected from previous queries. Currently, Constellation offers four built-in data models: LinearRegression, PolynomialRegression, Consistent, and Cyclic.\par
The linear and polynomial regression cache models offer simple yet widely applicable models for device data patterns. These models run a regression on the recently saved query results to predict future query requests.\par
The consistent cache model is the most trivial of the built-in models. It is expecting a data pattern of values that remain consistent with little variation from query to query. So, it simply predicts the current request to be the previous query result. This model can be used to support any data type, discussed further in the next section, while other models may be more limited (e.g. running a linear regression on images from a video will likely not generate a useful result).\par
The cyclic model is an example of a slightly more complex model that was designed to extend support for devices that display repeated patterns over a period of time. For example, this model may be appropriate for modelling device data that follows a diurnal pattern such as temperature or amount of sunlight. An Autoregressive Integrated Moving Average (ARIMA) model is used to model previous query results and predict future queries, similar to the prediction approach used in the regression-based models.
\subsection*{Extensibility}
To support the variety of data types and patterns collected by IoT devices, the edge node cache is designed to be extended to support new data types and models. Both extensions are designed to be fairly plug-n-play by implementing the corresponding Java interface and specifying the data type and model to be used in the device driver configuration file.\par
The cache can first be extended to support the data type that is being retrieved for the SENSE request. The interface only requires an implementation of a diff function which outputs the difference between two instances of the data type.\par

\begin{Verbatim}[fontsize=\small]
double diff(Object one, Object two)
\end{Verbatim}

Currently, Constellation offers three built-in data types: Double, CartesianCoordinates, and Image. The diff functions for these types return the absolute value of the difference, Euclidean distance between points, and sum of RGB pixel differences respectively.\par
The cache can also be extended by introducing a new data model. The built-in models described above provide a basis for simple data patterns, but it may be beneficial to create a new model that is more expressive for a specific device. The main functions present in the model interface are shown below.
\begin{Verbatim}[fontsize=\small]
// Add result to the model
void addPoint(int time, Object value)
// Return the expired time to predict
int getExpirationTime(double delta)
// Predict the value at the time specified
Object predictValue(int time)
\end{Verbatim}

Model extensions can be used not only to introduce new generally applicable models but also to create models designed for specific devices. For example, a thermometer device could introduce a model specifically designed for forecasting weather patterns. More accurate models result in improved prediction accuracy and higher query reduction rates.
\subsection*{Query Rate vs. Error Trade-off}
The cache query prediction mechanism for reducing the amount of device querying also introduces some amount of prediction error. Since applications will typically have different tolerances for error, the caching system provides configurations to control the trade-off between query reduction and error. This is done through the DELTA and ERROR values given in a CQL SENSE command.\par
The DELTA timeout value is used to ensure that the cache model does not extrapolate beyond a time when the data pattern becomes difficult to model and predict. The value specifies a hard requirement for when the cache must be bypassed resulting in a device query. For example, a DELTA timeout of 15 minutes says that the device must be queried if the current request is being made more than 15 minutes after the previous request. A smaller DELTA timeout enforces a tighter constraint resulting in lower prediction error at the expense of less query reduction. A larger DELTA timeout trades-off some additional prediction error for higher query reduction.\par
The ERROR value is used to stop cache prediction when recent predictions don't meet the specified error tolerance. This is done by comparing a recent query result with the corresponding cache prediction. For example, an ERROR value set at 2\textdegree C with a thermometer query result of 27\textdegree C while the cache predicted the temperature was only 24\textdegree C would result in cache prediction being temporarily disabled. Cache prediction is restarted when the cache model has accurately predicted a number of requests within the error tolerance. This constraint is especially important when the device is displaying volatile or unexpected data patterns that should be queried directly instead of predicted in the cache.
\subsection*{Evaluation}
The edge node caching system was evaluated using a collection of time series data simulated through a virtual device. The virtual device acts as a regular IoT device but uses an existing dataset for returning SENSE query results instead of an actual physical sensor. The time series dataset used is made up of a collection of meteorological data measures (including temperature, humidity, CO2 levels, etc.) \cite{edge_node_caching_dataset}.\par
The cache performance was measured in terms of two metrics: cache prediction error and query reduction rate. Cache prediction error measures the inaccuracy of cache hit prediction compared to the ground truth sensor value, which is available in the simulated dataset. Query reduction rate measures the percent of queries that are serviced by the cache and therefore do not need to be sent to the device.\par
Fig. \ref{cache_results} displays the performance results of the LinearRegression, PolynomialRegression, and Cyclic cache models for two temperature sensors. The query used for these tests was the following: SENSE Temperature FROM Thermometer DELTA 1 HRS ERROR 2 PERIOD 15 MINS. This CQL query requests a device value every 15 minutes and requires that the cache be bypassed at least once every hour (limiting the maximum query reduction rate to 1 - 15/60 = 75\%). Additionally, this query specifies that the cache should be bypassed if an error of 2\textdegree C is identified.\par

\begin{figure}[htbp]
\centerline{\includegraphics[width=\linewidth]{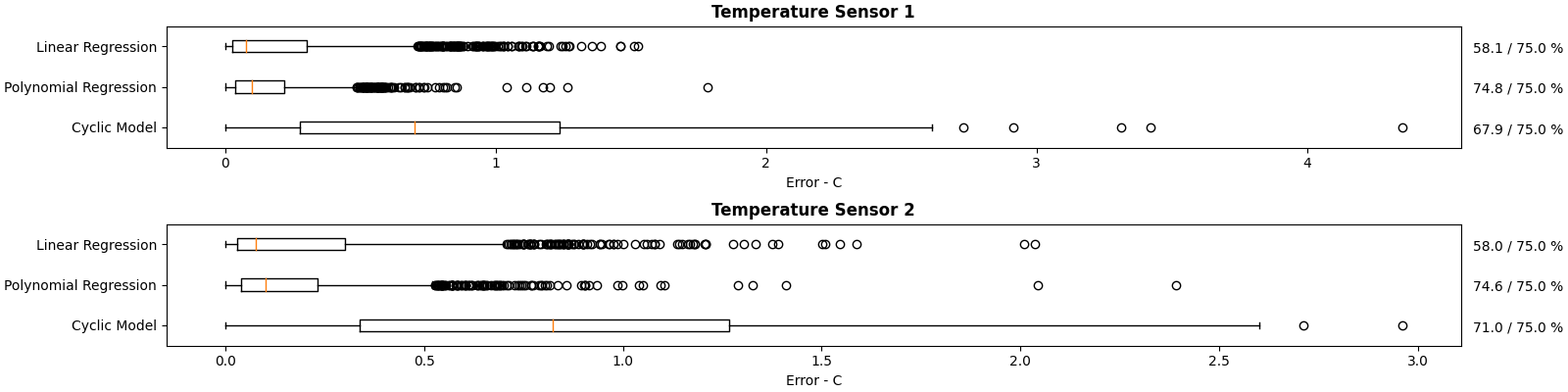}}
\caption{Query accuracy using various cache models for two different temperature sensors. Query reduction rate shown on the far right for each test (higher is better)}
\label{cache_results}
\end{figure}

For this query, the PolynomialRegression model is able to provide the best overall performance with a median prediction error of only about 0.1\textdegree C and a query reduction rate of about 74.7\%. Prediction errors generally remain low but reached up to 2.5\textdegree C when the temperature is quite volatile. Assuming the prediction error seen is within the application's tolerance, the system benefits from a significant reduction in the amount of querying.\par
As previously mentioned, the DELTA value can be used to trade-off prediction accuracy and query reduction rate. Fig. \ref{cache_query_vs_error} shows this trade-off for the temperature dataset.\par

\begin{figure}[htbp]
\centerline{\includegraphics[width=\linewidth]{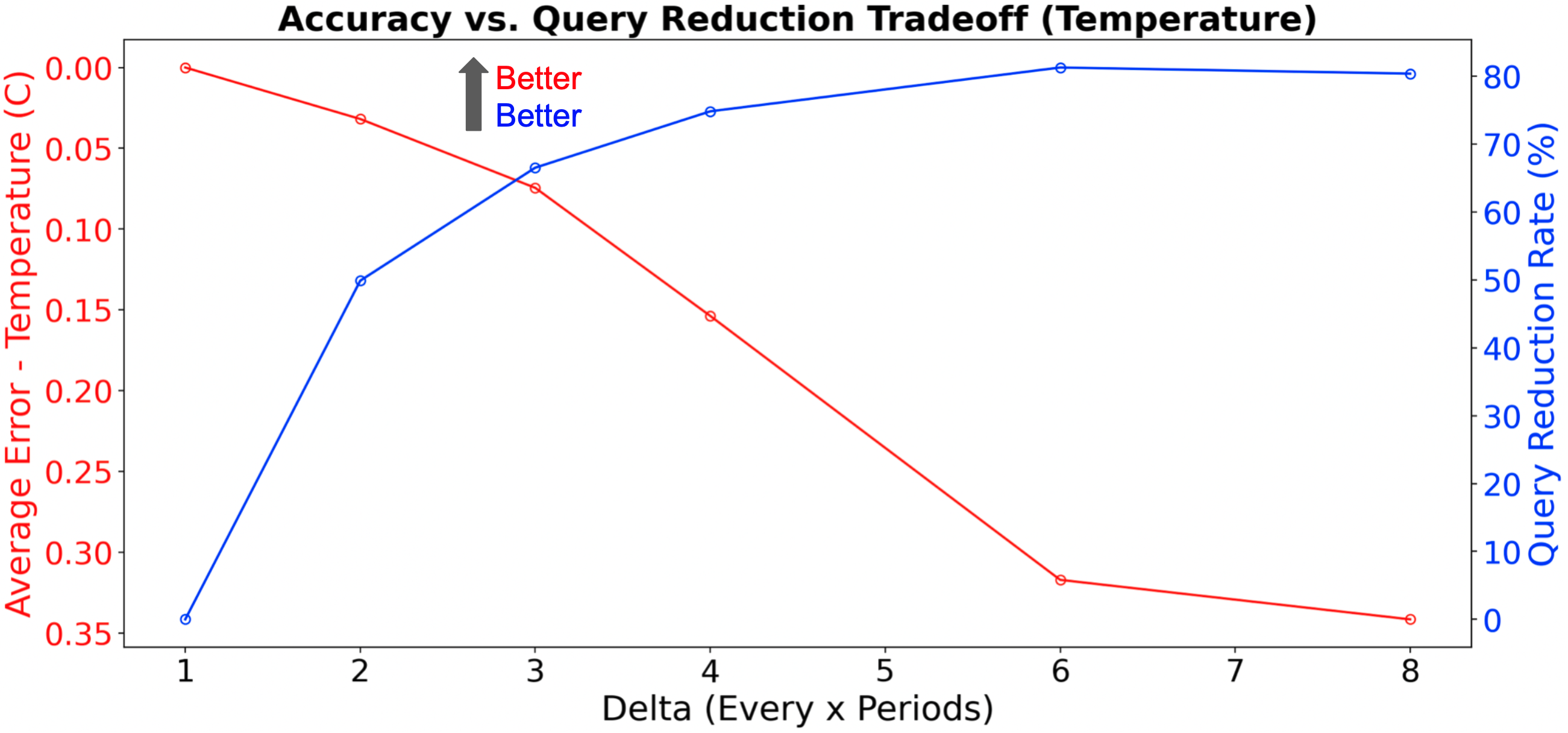}}
\caption{Prediction accuracy vs. query reduction rate trade-off for temperature dataset using polynomial regression model}
\label{cache_query_vs_error}
\end{figure}

If the client application has no tolerance for query error, then the DELTA timeout is not specified and the cache is never utilized (resulting in no query reduction). If the client application is willing to tolerate some error, then a large DELTA timeout can be provided to achieve a higher query reduction rate. Most applications will likely benefit from a DELTA value that balances the trade-off between these two extremes.\par

\section{Core 4: Data Privacy}
Constellation exhibits scalability, heterogeneity, high connectivity, and high data sharing. These attributes lead it to be exposed to threats and attacks. In particular, some IoT sensors, including healthcare devices and smart home devices, may contain significantly sensitive data, which raises privacy concerns from the end-users. In this section, the privacy and security requirements are specified with consideration of the trust model. Furthermore, we deployed data encryption, digital signature, and privacy mediators to protect the users’ data and provide the flexibility of data management.
    
\subsection*{Privacy Requirements}    
There are four main security and privacy requirements for Constellation: authenticity, integrity, confidentiality, and privacy. Authenticity is the critical problem of such a networked system. The core idea is to authorize entities as uniquely identifiable by a set of secure, non-replicable credentials. The integrity of messages can also be guaranteed through digital signatures. Confidentiality implies that messages and data should be accessible to only those who are authorized. Privacy requires protection mechanisms of secure sensitive data of both IoT devices and edge nodes. 

\subsection*{Privacy Mechanisms}    
In Constellation, we utilize digital signatures as unique identifications. Message owners sign the content of the message with its unique digital signature and which prevents unauthorized updates from data and messages. For confidentiality, encryption is a classic way to prevent unauthorized access from sensitive data. Constellation deploys ECDSA signature and ECC encryption because of the low latency with high security.\par
For data privacy, one of the main privacy concerns is data leakage caused by third-party applications. There are a variety of reasons for data leakage. The applications may have various security guarantees and mechanisms, the end-users are not complex enough to handle such privacy policies. In addition, the service provider may be hacked or sell data to malicious people. Moreover, IoT devices are ubiquitous and more likely to contain private information; therefore, granting hundreds of applications and clients is not safe practice.\par
Based on these observations, the privacy mediator is introduced to allow the users to “control the release of their data”\cite{davies2016privacy}. The purpose of the mediator is to allow the users to develop their privacy policy individual sensors rather than handling this privacy on an application to application basis. To be compatible with the devices produced by various manufacturers and working with different device drivers, the mediator serves the sensors’ SENSE queries to edge nodes and enforces user's privacy preferences by pre-processing the data before releasing to the clients and applications. It also checks the identity of clients and encrypts prepared data to prevent authentication attacks that seek to perform unauthorized accesses. In this model, the users place their trust on the edge nodes and mediators that serve their sensors.\par
The sensor owners can develop the privacy policy towards its sensors and the mediator enforces the given privacy policies. Originally, the privacy mediator provides three types of privacy controls: delete, denature, and summarize. The queries to define the privacy policy are as follows.
\begin{Verbatim}[fontsize=\small]
// Three rules: delete, denature, summarize 
denature sensor [sensorID] [rules] 
\end{Verbatim}

Specifically, deletion means blocking the access and hiding the data from the certain clients. Under the deletion policy, the sensor owners can authorize clients by developing the whitelist or blocklist of the clients. In particular, the sensor owner can specify whether a type of data can be accessed by a specific client. For instance, the sensor owner can set a rule in the mediator to block an application from accessing the temperature of a sensor. In addition to simply blocking, It also provides more flexible denaturing algorithms. Denaturing is to modify the sensor data before releasing. Instead of simply blocking, denaturing allows the owners to take more operations to alleviate the potential leakage. For example, the mediator can replace the original data with a given text or blur the original data by inserting randomly generated characters. The denaturing operation is important when the sensor contains images, videos, and audios. Summarizing is used to encapsulate the original data and send an opaque report to the client. For example, the constrained clients and applications can only obtain the zip code instead of the specific address of the sensor. Similarly, the user can choose to share the average when handling temperature-related data requests. \par

The privacy mediator also provides a Java-based interface and the owners are able to include more personalized mediation algorithms.\par
\begin{Verbatim}[fontsize=\small]
// denature interface
object Denature(
String sensorID, object originalData);
// summarize interface
object Summarize(
String deviceID, object originalData);
\end{Verbatim}

\subsection*{Evaluation}
Constellation provides real-time responses. Thus, overheads of privacy mechanisms are the main concerns in terms of evaluation. \par
The previous work, such as \cite{dhanda2020lightweight,suarez2018practical} have researched and analyzed various cryptography methods in terms of overheads and security levels. Both hardware and software efficiency of ECC cryptography is evaluated in this work. Similarly, \cite{toradmalle2018prominence} has assessed the performance of digital signature schemes. ECDSA is chosen because of its high-security level and limited resources of IoT devices. \par
For the privacy mediator, the three built-in control rules are evaluated. Deletion is directly removing the sensitive data before releasing. Denaturing is inserting random characters in the original text to blur the data. Summarizing is assessed for physical address data. In this experiment, it generates the corresponding zip code and hides the actual address from the clients and applications. \par

\begin{figure}[htbp]
\centerline{\includegraphics[scale=0.4]{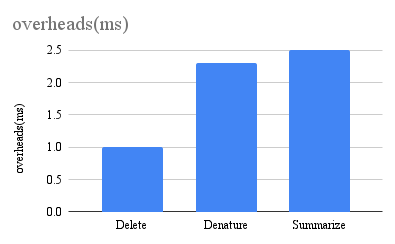}}
\caption{Overheads of three built-in privacy policy rules}
\label{priv_results}
\end{figure}
The experimental results demonstrate that the privacy mediators are able to enforce various privacy policies with low overheads and high security.

\section{Conclusion}
To support complicated applications in heterogeneous IoT environments, an edge-based semantic runtime system is proposed. One of main concerns is to address the programming complexity and provide an unified view of tasks management in the edge perspective. Specifically, This work demonstrates four ideas to support fault-tolerance, scalablity, caching, and privacy. 

\section*{Acknowledgment}
This research was supported in part by an NSF grant to the Distributed Computing Systems Group at the University of Minnesota (CNS-1908566). 

\bibliographystyle{IEEEtran}
\bibliography{bib}
\vspace{12pt}

\end{document}